\documentstyle[12pt]{article}
\begin{document}
\pagestyle{empty}
\hspace*{12.4cm}IU-MSTP/48 \\
\hspace*{13cm}hep-th/0206232 \\
\hspace*{13cm}June, 2002
\begin{center}
 {\Large\bf Toward Construction of Exact Operator Solution 
   of $A_N$-Toda Field Theory}
\end{center}

\vspace*{1cm}
\def\thefootnote{\fnsymbol{footnote}}
\begin{center}{\sc Y. Takimoto}$^1$ and {\sc T. Fujiwara}$^2$ 
\end{center}
\vspace*{0.2cm}
\begin{center}
{\em $\ ^{1}$ Information Processing Center, Ibaraki University, Mito 310-8512, Japan}\\
{\em $\ ^{2}$ Department of Mathematical Sciences, Ibaraki University,
Mito 310-8512, Japan}\\
\end{center}
\vfill

\begin{center}
{\large\sc Abstract}
\end{center}

Quantum $A_N$-Toda field theory in two dimensions is investigated 
based on the method of quantizing canonical free field. Toda 
exponential operator associated with the fundamental weight 
$\lambda^1$ is constructed. 

\vskip .3cm

\newpage

\pagestyle{plain}
Toda field theories (TFT) played prominent roles as completely integrable 
two dimensional field theory in various branches of physics  
from statistical field theory to string physics. In classical theory
exact solutions are known \cite{ls} for TFT associated with 
arbitrary simple Lie algebra, and many interesting aspects have been 
investigated so far \cite{Babe,bg,bg89,bfo,wgeom}. 

The simplest TFT associated with the Lie algebra $A_1$ is 
the Liouville theory, for which exact operator solutions have been 
obtained by several authors \cite{bcgt,gn,ow,kn,gs93,fit96}. 
Bilal and Gervais \cite{bg, bg89} have investigated quantization of 
TFT based on the extensive canonical analysis developed for the Lioville 
theory \cite{ct,gn84}. The full operator solution for $A_2$ case has also been obtained in 
ref. \cite{fit98,tikf}. However, the generalization of such exact 
operator solutions to an arbitrary TFT is still lacking. In this note, 
we investigate the construction of the Toda exponential operator 
associated with the fundamental weight $\lambda^1$ for an arbitrary 
$N$ as a first step toward the complete construction of higher rank 
solutions. 

Let us begin with the classical action of TFT associated with an 
arbitrary rank $N$ simple Lie algebra 
\begin{eqnarray}
\label{action}
  S=\frac{1}{\gamma^2}\int_{-\infty}^{+\infty}d\tau
  \int_0^{2\pi}d\sigma \Biggl(\frac{1}{2}\partial_\alpha\varphi\cdot
  \partial^\alpha\varphi-\mu^2\sum_{a=1}^N
  {\rm e}^{\alpha^a\cdot\varphi}\Biggr) ~,
\end{eqnarray}
where $\varphi$ is the $N$-component Toda fields, and we 
work in the Cartan-Weyl basis. $\alpha^a$ $(a=1,\cdots ,N)$ are the 
simple roots, and $\lambda^a$ are the fundamental weight vectors 
satisfying $2\lambda^a\cdot\alpha^b/(\alpha^b)^2=\delta^{ab}$. 
In spite of the dimension full mass parameter $\mu^2$ the action 
(\ref{action}) possesses the conformal invariance. At the classical 
level the coupling constant $\gamma$ is a free parameter. In quantum 
theory it must be chosen by the requirement of conformal invariance 
in the presence of conformal matters \cite{itf98}. 
The classical Toda field equations 
\begin{eqnarray}
\label{eqofmotion}
  \partial_\alpha\partial^\alpha\varphi-\mu^2\sum_{a=1}^N \alpha^a
  {\rm e}^{\alpha^a\cdot \varphi}=0
\end{eqnarray}
can be solved exactly by Lie algebraic method \cite{ls,Babe}. 

To obtain complete operator solution we need all the Toda 
exponential operators for arbitrary fundamental weights \cite{fit98,tikf}. 
In what follows, however, we will construct the exponential 
operator associated with $\lambda^1$ only. This restriction is 
mainly due to the reason that the exchange algebra among the 
screening charges appearing in the expansion of the operator 
associated with other weight vectors become very complicated. 

The classical Toda exponential associated with $\lambda ^1$ 
is explicitly given by 
\begin{eqnarray}
  \label{clexp}
  {\rm e}^{\lambda^1 \cdot \varphi (\tau ,~\sigma ) }
  =\frac{{\rm e}^{\lambda^1 \cdot 
      \psi (\tau ,~\sigma ) }}{1
    +\sum_{n=1}^{N}\left(\frac{\mu^2}{4}\right)^n 
    A_{1\cdots n}(x^+)B_{1\cdots n}(x^-)}
\end{eqnarray}
where $\psi(x)=\psi_+(x^+)+\psi_-(x^-)$ with $\psi_\pm(x^\pm)$ being 
arbitrary functions of the light-cone variables $x^\pm=\tau\pm\sigma$. 
It is identified with the canonical free field. The chiral screening 
charges $A_{1\cdots n}(x^+)$ and $B_{1\cdots n}(x^-)$ are defined by the 
differential equations
\begin{eqnarray}
\label{dela}
  \partial_+A_1(x^+)=V_1^+(x^+)~, \qquad 
  \partial_+A_{1\cdots n}(x^+)=V_1^+(x^+)A_{2\cdots n}(x^+)~,
\end{eqnarray}
where $V^\pm_a(x^\pm)={\rm e}^{\alpha^a\cdot\psi_\pm(x^\pm)}$. 

To ensure the periodic boundary condition on $\varphi$, we assume the 
same periodicity for $\psi$. The left- and the right-moving modes 
$\psi_\pm$, however, are not $2\pi$ periodic. To see this, we note the 
normal mode expansions 
\begin{eqnarray}
  \psi_\pm(x^\pm)=\frac{\gamma}{2}Q+\frac{\gamma}{4\pi}Px^\pm
  +\frac{i\gamma}{\sqrt{4\pi}}\sum_{n\neq 0}\frac{1}{n}a_n^{(\pm)}
  {\rm e}^{-inx^\pm} ~.
\end{eqnarray}
Then $\psi_\pm$ satisfy $\psi(x^\pm\pm2\pi)
=\psi_\pm(x^\pm)\pm\displaystyle{\frac{\gamma}{2}P}$. 
The chiral screening charges defined by (\ref{dela}) are given by 
\begin{eqnarray}
\label{clchirala}
A_{1\cdots n}(x)\hskip -.2cm&=&\hskip -.2cm 
  C_{\alpha^1+\cdots+\alpha^n}
  C_{\alpha^2+\cdots+\alpha^n}\cdots C_{\alpha^n}
  \int_0^{2\pi}\hskip -.15cm dy_1
  {\cal E}_{\alpha^1+\cdots+\alpha^n}(x-y_1) 
  V^+_1(y_1) \nonumber\\  
  \hskip -.2cm&&\hskip -.2cm \times \int_0^{2\pi}\hskip -.15cm dy_2
  {\cal E}_{\alpha^2+\cdots+\alpha^n}(y_1-y_2)V^+_2(y_2)
  \cdots\int_0^{2\pi\hskip -.15cm }dy_n{\cal E}_{\alpha^n}(y_{n-1}-y_n)
  V^+_n(y_n)~. 
\end{eqnarray}
The periodicity of $\varphi(x)$ is guaranteed by the periodicity of 
$A(x^+)B(x^-)$. 
In (\ref{clchirala}) we have introduced for an arbitrary vector 
$\beta$ in the root space
\begin{eqnarray}
C_\beta=\Biggl(2{\rm sinh}\frac{\gamma}{4}\beta\cdot P\Biggr)^{-1}~, \qquad
  {\cal E}_\beta(x)=\exp \frac{\gamma}{4} \beta\cdot P \epsilon(x)
\end{eqnarray}
with $\epsilon(x)$ being the stair-step function defined by 
$\epsilon(x)=1$ for $0<x<2\pi$ and $\epsilon(x+2\pi)=\epsilon(x)+2$.
The screening charges (\ref{clchirala}) satisfy the quasi-periodicity
\begin{eqnarray}
\label{clperio}
  A_{1\cdots n}(x^++2\pi)={\rm e}^{\frac{\gamma}{2}(\alpha^{1}
    +\cdots+\alpha^{n})\cdot P}A_{1\cdots n}(x^+) ~.
\end{eqnarray}
The screening charges in the right-moving sector $B_{1\cdots n}$ 
are obtained by replacing $V_a^+$ with $V_a^-$ in the rhs of (\ref{clchirala}).

A remarkable feature of TFT is that the mapping from $\varphi$ to 
$\psi$ described by the classical solution preserves the 
canonical structure \cite{Babe}. In other words the 
fundamental Poisson brackets between the canonical variables $\varphi$ 
and $\pi_\varphi\equiv\displaystyle{\frac{1}{\gamma^2}\partial_\tau\varphi}$
are guaranteed by the canonical pairs of the free fields  $\psi$ and 
$\pi_\psi\equiv\displaystyle{\frac{1}{\gamma^2}\partial_\tau\psi}$, 
or equivalently, by the Poisson brackets
\begin{eqnarray}
  \{\psi_{k\pm}(x^\pm),\psi_{l\pm}({x'}^\pm)\}=\frac{\gamma^2}{4}
  \Biggl(\frac{x -x'}{2\pi}-\epsilon(x^\pm-{x'}^\pm)\Biggr)\delta_{kl}~, 
\end{eqnarray}
where the indices $k,l=1,\cdots ,N$ stand for the components of the free 
field $\psi$ \cite{tikf}. Furthermore, the theory possesses extended 
conformal invariance, whose generators reduce to the free field expressions 
after the substitution of the classical solution obtained from (\ref{clexp}). 
In particular, the stress tensor generates the 
pseudo-conformal symmetry for the canonical free fields exactly the same 
manner as for the interacting fields. The virtue in writing the classical 
solution in the form (\ref{clexp}) is that the conformal property becomes  
manifest. Since both the interacting and the free classical vertex 
functions possess the same transformation properties under the conformal 
symmetry, the $N$ screening charges $A_{1\cdots n}(x^+)B_{1\cdots n}(x^-)$  ($n=1,\cdots ,N$) 
must be of vanishing conformal weight. 
This can be verified directly from (\ref{clchirala}).

As was argued in detail in ref. \cite{tikf}, it is convenient to make the 
substitutions 
$Q\rightarrow 2Q$ in $\psi_\pm(x^\pm)$ and introduce the 
$\star$-product between $L$ depending only on $\psi_+$ and $R$ 
depending only on $\psi_-$ by 
\begin{eqnarray}
  \label{lr}
  L\star R\equiv L{\rm e}^{-\gamma\omega\cdot Q}R~, 
\end{eqnarray}
where $L$ and $R$ are assumed to have the same dependences on $Q$ and 
$\omega$ is chosen to satisfy $\{L{\rm e}^{-\gamma\omega\cdot Q},
{\rm e}^{-\gamma\omega\cdot Q}R\}=0$. 
This not only satisfies the product rule
\begin{eqnarray}
  \label{prodrulr}
  L\star R L'\star R'=LL'\star RR'~,
\end{eqnarray}
but also preserve the chiral structure
\begin{eqnarray}
  \label{eq:cpbr}
  \{L\star R,L'\star R'\}
  =\{L,L'\}\star RR'+LL'\star\{R,R'\}~.
\end{eqnarray}

We now turn to quantum theory. We impose the standard commutation relations 
on the normal modes
\begin{eqnarray}
  \label{comuqp}
  [Q_k,P_l]=i\delta_{kl} ~, \qquad [a_{kn}^{(+)},a_{lm}^{(+)}]
  =[a_{kn}^{(-)},a_{lm}^{(-)}]=n\delta_{kl}\delta_{n+m,0} ~,\qquad (k,~l=1,\cdots ,N)~.
\end{eqnarray}
The chiral vertex operators are defined by the normal ordering for the 
oscillator modes and the symmetric ordering 
for the zero-modes $P$, $Q$ by the rule $:{\rm e}^{\beta\cdot Q}f(P):
={\rm e}^{\frac{1}{2}\beta\cdot Q}f(P){\rm e}^{\frac{1}{2}\beta\cdot Q}$. 
We write chiral screening operators using simple notation
\begin{eqnarray}
 \label{quana1}
  A_{\{n\}}(x)&=&\int_0^{2\pi}dy_1dy_2\cdots dy_n
  :{\cal E}_{\alpha^1}(x-y_1)V^+_1(y_1):
  :{\cal E}_{\alpha^2}(x-y_1)
  {\cal E}_{\alpha^2}(y_1-y_2)V^+_2(y_2):\nonumber\\
  &&\times\cdots\times
  :{\cal E}_{\alpha^n}(x-y_1){\cal E}_{\alpha^n}(y_1-y_2) 
  \cdots {\cal E}_{\alpha^n}(y_{n-1}-y_n)V^+_n(y_n):~,
\end{eqnarray}
where $\{n\}$ stands for $1\cdots n$ and we employ the convention 
$A_{\{0\}}=B_{\{0\}}=1$ for later convenience. 
Two major modifications are made here from the classical expressions 
(\ref{clchirala}). The one is the omission of the overall $P$ dependent 
coefficients to simplify the operator algebra satisfied 
by the screening charges. The other is the rescaling 
$\psi_+\rightarrow \eta\psi_+$ to keep the conformal weight of the vertex 
operator $V^+_a$ to be $(1,0)$ \cite{ow,ct,gn84}. 
Then the conformal weights of the screening charges vanish. 
We assume that the conformal symmetry is generated by the normal-ordered 
free field stress tensor
\begin{eqnarray}
  T_{\pm\pm}&=&\frac{1}{\gamma^2}
  (:\partial_\pm\psi\cdot\partial_\pm\psi:
  -2\rho\cdot\partial_\pm^2\psi) ~,
\end{eqnarray}
where $\rho$ is assumed to satisfy $\alpha^a\cdot\rho=1$ for any simple root 
and is explicitly given by $\rho=\sum_a\frac{2\lambda^a}{(\alpha^a)^2}$. 
The conformal weight $\Delta_\beta$ of the vertex operators
$:{\rm e}^{\eta\beta\cdot\psi}:$ is 
\begin{eqnarray}
  \label{cdim}
  \Delta_\beta=\beta\cdot\Bigl(\eta\rho-\frac{\gamma^2\eta^2}{8\pi}
  \beta\Biggr)~.
\end{eqnarray}
In particular $\eta$ must satisfy 
\begin{eqnarray}
  \label{ita}
  \Delta_{\alpha^a}=\eta-\frac{\gamma^2\eta^2}{4\pi}=1 ~.
\end{eqnarray}
We also make the replacement  $P$ by $\eta P$ in 
${\cal E}_\beta(x)$ to ensure the quasi-periodicity of the screening 
charges under the shift $x\rightarrow x+2\pi$. 
The screening charges $B_{\{n\}}$ in the right-moving sector 
are similarly defined. 

The $\star$-product (\ref{lr}) can be defined also in quantum theory. It 
satisfies the product rule (\ref{prodrulr}), which implies that the left- 
and the right-chiral operators can be considered as commuting under the 
$\star$-product. In particular one easily see $L\star f(\varpi)R
=f(\varpi)L\star R$ for an arbitrary $f$ as a special case. For later 
convenience, we introduce here some notation
\begin{eqnarray}
  \label{somedef}
  \varpi\equiv-\frac{iP}{\gamma\eta}~, \qquad
  \varpi^a\equiv\alpha^a\cdot\varpi ~, \qquad
  g\equiv\frac{\gamma^2\eta^2}{8\pi}=\frac{\eta-1}{2}~, \qquad
  q\equiv{\rm e}^{2\pi ig}~, 
\end{eqnarray}
where $q$ is the quantum deformation parameter and $g$ plays the role 
of the Planck constant. A $q$-number $[x]$ is defined as usual by
\begin{eqnarray}
  \label{eq:qn}
  [x]=\frac{q^x-q^{-x}}{q-q^{-1}}~.
\end{eqnarray}
We also use $[x]_n\equiv [x][x+1]\cdots[x+n-1]$ and $[n]!\equiv
[1][2]\cdots[n]$.
In terms of these 
variables (\ref{quana1}) can be written as
\begin{eqnarray}
  \label{eq:sc}
  A_{\{n\}}(x)&=&\int_0^{2\pi}dy_1dy_2\cdots dy_n
    q^{(\Omega_n+1)\epsilon(x-y_1)
    +(\Omega_n -\Omega_1)\epsilon(y_1-y_2) +\cdots (\Omega_{n} -\Omega_{n-1})
    \epsilon(y_{n-1}-y_n)}\nonumber\\
    &&\hskip 3cm \times V_1^+(y_1)V^+_2(y_2)\cdots V^+_n(y_n)~,
\end{eqnarray}
where we have introduced $\Omega_n$ by 
\begin{eqnarray}
\Omega_n \equiv \sum_{a=1}^{n}\varpi^a~,\qquad\Omega_0 \equiv 0~.
\end{eqnarray}
In quantum theory the quasi-periodicity (\ref{clperio}) is modified to
\begin{eqnarray}
  \label{quanperio}
  A_{\{n\}}(x+2\pi)=q^{2(\Omega_n+1)}A_{\{n\}}(x) ~.
  \end{eqnarray}

The screening charges (\ref{eq:sc}) and the vertex operators 
\begin{eqnarray}
  \label{eq:vopk}
  V^1_\kappa(x)\equiv V^{1+}_\kappa(x^+)\star V^{1-}_\kappa(x^-)
  \quad \hbox{with} \quad V^{1\pm}_\kappa(x^\pm)
  \equiv :{\rm e}^{\kappa\eta\lambda^1\cdot\psi_\pm(x^\pm)}:
\end{eqnarray}
are the building blocks of the Toda exponential operators 
$e^{\kappa\eta\lambda^1\cdot\varphi}$. These 
satisfy mutual commutativity among $V^1_\kappa (x)$ and $A_{\{n\}}(x)$ 
at the same point,\footnote{The commutativity of the screening charges 
$A_{\{n\}}(x)A_{\{m\}}(x)=A_{\{m\}}(x)A_{\{n\}}(x)$ is simply assumed 
here. This can be shown explicitly in the $A_2$-case. At the time writing 
this paper, the satisfactory proof of this is not known to the present 
authors.} and the conformal weight of the screening charges vanishes. 
We thus expand the Toda exponential as 
\begin{eqnarray}
  \label{expkappa}
  {\rm e}^{\kappa\eta\lambda^1\cdot\varphi(x)}
  =V_\kappa^1(x)\sum_{n_1,\cdots  ,n_N=0}^{\infty}\Biggl(\frac{\mu^2}{4}
\Biggr)^{\sum_{l=1}^{N}ln_l}C_{n_1 \cdots n_N}^1(\kappa;\varpi) \prod_{m=1}^N
(A_{\{m\}}(x^+)\star B_{\{m\}}(x^-))^{n_m}~,
\end{eqnarray}
where $C^1_{n_1 \cdots n_N}$ are unknown coefficients depending on the 
zero-mode momenta without affecting the conformal property. We assume
\begin{eqnarray}
  C_{0\cdots 0}^1(\kappa;\varpi)=1 ~, \quad C^1_{n_1\cdots n_N}(0;\varpi)
  =\delta_{n_10}\cdots
  \delta_{n_N0}~.
\label{cn}
\end{eqnarray}
Note that there arises no ordering ambiguity due to the assumed 
commutativity of operators mentioned above. 

At this point we give an outline of the strategy for the construction of 
(\ref{expkappa}). The first step is to construct the exponential operator 
for $\kappa =-1$
\begin{eqnarray}
{\rm e}^{-\eta \lambda^1 \cdot \varphi(x)}&=&V_{-1}^1(x)\sum_{n=0}^{N}
\Biggl(\frac{\mu^2}{4}\Biggr)^{n}C^1_{0\cdots010\cdots0}(-1;\varpi)
A_{\{n\}}(x^+)\star B_{\{n\}}(x^-) \nonumber \\
&=&\sum_{n=0}^{N}\Biggl(\frac{\mu^2}{4}\Biggr)^{n}C^1_{[n]}(\varpi)
\Psi^{1+}_{\{n\}}(x^+)\star\Psi^{1-}_{\{n\}}(x^-)
\label{inv}~,
\end{eqnarray}
where $C^1_{[n]}(\varpi)$ is given by 
\begin{eqnarray}
  \label{eq:c1n}
C^1_{[n]}(\varpi)=C^1_{{\scriptsize\underbrace{0\cdots 0}_{n-1}\mbox{1}
\underbrace{0\cdots 0}_{N-n}}}(-1;\varpi-\lambda^1)~, 
\end{eqnarray}
and $\Psi^{1\pm}_{\{n\}}(x^\pm)$ are the chiral fields defined by
\begin{eqnarray}
  \label{eq:cf}
  \Psi_{\{n\}}^{1+}(x^+)\equiv V^{1+}_{-1}(x^+)A_{\{n\}}(x^+)~, \qquad
  \Psi_{\{n\}}^{1-}(x^-)\equiv V^{1-}_{-1}(x^-)B_{\{n\}}(x^-)~.
\end{eqnarray}
The coefficients $C^1_{[n]}$ must be chosen to satisfy the locality 
condition
\begin{eqnarray}
  [{\rm e}^{-\eta\lambda^1\cdot\varphi(0,\sigma)},
{\rm e}^{-\eta\lambda^1\cdot\varphi(0,\sigma')}]=0~.
\label{locality}
\end{eqnarray}
Once we have the operator for $\kappa=-1$, we can construct (\ref{expkappa}) 
for an arbitrary negative integer $\kappa$ as a composite operator. In this 
way we are able to find the expansion coefficients 
$C_{n_1 \cdots n_N}^1(\kappa;\varpi)$ for an arbitrary negative integer 
$\kappa$. This is the second step. We finally continue $\kappa$ to an 
arbitrary real number. 

To construct the operator (\ref{inv}) we need the exchange algebra among 
$V^{1+}_\kappa (x)$ and $A_{\{n\}}(x)$, which can be inferred from the 
results of the $A_2$-case \cite{tikf}. It is straightforward to find the 
exchange algebra between $V^{1+}_\kappa (x)$ and $A_{\{n\}}(x)$. As for 
the screening charges, we assume the general form
\begin{eqnarray}
  \label{eq:exchalg}
  A_{\{n\}}(x)A_{\{m\}}(x')&=&\alpha_{nm}A_{\{m\}}(x') A_{\{n\}}(x)
  +\beta_{nm}A_{\{n\}}(x') A_{\{m\}}(x) \nonumber\\
  &&+\gamma_{nm}A_{\{n\}}(x)A_{\{m\}}(x)+\delta_{nm}A_{\{n\}}(x')A_{\{m\}}(x')~.
\end{eqnarray}
The coefficients appearing in the rhs' of (\ref{eq:exchalg}) 
depend on $\varpi$ 
as well as the sign of $x-x'$. They can be obtained straightforwardly 
by applying the method 
developed in ref. \cite{tikf}. We thus find the exchange algebra
\begin{eqnarray}
\label{av}
V^{1+}_\kappa(x)V^{1+}_\nu(x')&=&q^{{-\kappa\nu}(\lambda^1)^2\epsilon(x-x')}
V^{1+}_\nu(x')V^{1+}_\kappa(x) \nonumber\\
A_{\{n\}}(x)V^{1+}_\kappa(x') 
&=&\frac{q^{-\kappa\epsilon (x-x')}[\Omega_n+1]}{[\Omega_n+\kappa+1]}
V^{1+}_\kappa(x') A_{\{n\}}(x) + 
\frac{q^{(\Omega_n+1)\epsilon (x-x')}[\kappa]}{[\Omega_n+\kappa+1]}
V^{1+}_\kappa(x') A_{\{n\}}(x')~, \nonumber\\
A_{\{n\}}(x)A_{\{m\}}(x')&=& 
-\frac{q^{(\Omega_n-\Omega_m-1)\epsilon (x-x')}}{[\Omega_n-\Omega_m]}
A_{\{n\}}(x')A_{\{m\}}(x) \nonumber\\
&&+\frac{q^{-\epsilon (x-x')}[\Omega_n+1][\Omega_n-\Omega_m-1]
[\Omega_m+2]}{[\Omega_n+2][\Omega_n-\Omega_m][\Omega_m+1]}A_{\{m\}}(x')
A_{\{n\}}(x)\nonumber \\
&&-\frac{q^{-(\Omega_m+2)\epsilon (x-x')}}{[\Omega_m+1]}
A_{\{n\}}(x)
A_{\{m\}}(x) \nonumber \\
&&+\frac{q^{(\Omega_n+1)\epsilon (x-x')}}{[\Omega_n+2]}
A_{\{n\}}(x')
A_{\{m\}}(x')~,\quad(n\neq m)\\
A_{\{n\}}(x)A_{\{n\}}(x') &=&
q^{-2\epsilon (x-x')}A_{\{n\}}(x')
A_{\{n\}}(x)-\frac{q^{-(\Omega_n+3)\epsilon (x-x')}}{[\Omega_n+2]}
(A_{\{n\}}(x))^2 \nonumber\\
&&+\frac{q^{(\Omega_n+1)\epsilon (x-x')}}{[\Omega_n+2]}
(A_{\{n\}}(x'))^2~. \nonumber
\end{eqnarray}
The exchange algebra among the chiral field $\Psi_{\{n\}}^{1+}(x)$ can also 
be obtained from (\ref{av}) as
\begin{eqnarray}
\Psi_{\{n\}}^{1+}(x)\Psi_{\{m\}}^{1+}(x')&=&\sum_{k,l}{\cal R}_{nm}^{lk}
\Psi_{\{k\}}^{1+}(x')\Psi_{\{l\}}^{1+}(x)~.
\label{bloch}
\end{eqnarray}
where ${\cal R}_{nm}^{kl}$ is the $\varpi $ dependent ${\cal R}$-matrix 
and the nonvanishing matrix elements are explicitly given by 
\begin{eqnarray}
  \label{Rmat}
  {\cal R}^{nn}_{nn}&=&q^{-\frac{2}{N-1}\epsilon(x-x')}~, \nonumber\\
  {\cal R}^{nm}_{nm}&=&q^{\frac{N-1}{N+1}\epsilon(x-x')}
  \frac{[\Omega_n-\Omega_m+1]}{[\Omega_n-\Omega_m]}~, \\
  {\cal R}^{mn}_{nm}&=&-\frac{q^{\bigl(\frac{N-1}{N+1}+\Omega_n-\Omega_m\bigr)
      \epsilon(x-x')}}{[\Omega_n-\Omega_m]}~. \quad (n\ne m)\nonumber
\end{eqnarray}
Essentially equivalent exchange algebra with (\ref{bloch}) has already been 
obtained in ref. \cite{bg}.  

We are now in a position to analyze the locality condition (\ref{locality}). 
Using the exchange algebra (\ref{bloch}) and the relation 
\begin{eqnarray}
  \label{eq:exchpi}
  \Psi_{\{n\}}^{1+}(x)\varpi=(\varpi-\mu_n)\Psi_{\{n\}}^{1+}(x)
\end{eqnarray}
with $\mu_n\equiv \lambda^1-\alpha^1-\cdots-\alpha^n$, we can examine the 
locality as
\begin{eqnarray}
  \label{eq:loc}
  0&=&[e^{-\eta\lambda^1\cdot\varphi(0,\sigma)},e^{-\eta\lambda^1\cdot\varphi(0,\sigma')}] \nonumber \\
  &=&\sum_{k,l}\Biggl\{\sum_{n,m}\sum_{r,s}C^1_{[n]}(\varpi)C^1_{[m]}(\varpi-\mu_n)
  {\cal R}^{kl}_{nm}(\varpi)
  \overline{\cal R}^{rs}_{nm}(\varpi)
  -C^1_{[l]}(\varpi)C^1_{[k]}(\varpi-\mu_l)\Biggr\} \nonumber\\
  &&\times\Psi^{1+}_{\{l\}}(\sigma')\Psi^{1+}_{\{k\}}(\sigma)
  \star\Psi^{1-}_{\{s\}}(-\sigma')\Psi^{1-}_{\{r\}}(-\sigma)~,
\end{eqnarray}
where $\overline{\cal R}^{kl}_{nm}(\varpi)$ is the inverse of 
${\cal R}^{kl}_{nm}$ defined by 
\begin{eqnarray}
  \label{eq:invR}
  \sum_{r,s}{\cal R}_{nm}^{rs}\overline{\cal R}_{sr}^{kl}
  &=&\delta_n^l\delta_m^k~.
\end{eqnarray}
It is now straightforward to show that the locality condition is 
equivalent to 
\begin{eqnarray}
  \label{eq:eqloc}
  C^1_{[k]}(\varpi)C^1_{[l]}(\varpi-\mu_k){\cal R}^{nm}_{kl}(\varpi)
  =C^1_{[m]}(\varpi)C^1_{[n]}(\varpi-\mu_m){\cal R}^{lk}_{mn}(\varpi)~.
\end{eqnarray}
Similar relation has already been obtained in ref. \cite{tikf}. 
Using the explicit expression (\ref{Rmat}), we obtain the constraints on 
$C^1_{[n]}(\varpi)$ as
\begin{eqnarray}
  \frac{C^1_{[n]}(\varpi)C^1_{[m]}(\varpi-{\mu}_n)}{C^1_{[n]}(\varpi-{\mu}_m)
  C^1_{[m]}(\varpi)}
=\frac{[\Omega_n -\Omega_m -1]}{[\Omega_n -\Omega_m +1]}~.
\label{cccc}
\end{eqnarray}
Unfortunately, these do not uniquely determine $C_{[n]}^1(\varpi)$. To proceed 
further we assume that the $\varpi$ dependences of $C_{[n]}^1(\varpi)$ are 
minimal and correctly reproduce the classical limit \cite{tikf}. Thus we are 
lead to the ansatz
\begin{eqnarray}
C^1_{[n]}(\varpi)
=c_n\prod_{k=0}^{n-1}
\frac{1}{[\Omega_{k}-\Omega_n+a_k][\Omega_{k}-\Omega_n+b_k]}~,
\label{acn}
\end{eqnarray}
where the unknown constants $a_k$ and $b_k$ are chosen so as for 
$C^1_{[n]}(\varpi)$ to satisfy (\ref{cccc}). It is easy to show 
that (\ref{cccc}) are satisfied by 
\begin{eqnarray}
a_k =0~,\qquad b_k =1~.
\label{unknowncnstab}
\end{eqnarray}
This result is completely consistent with the operator solutions for 
the Louville and the $A_2$ cases. The overall constant $c_n$ cannot 
be determined from the locality constraints (\ref{cccc}). It might be 
fixed by the requirement of the field equations \cite{fit96,fit98,tikf}. 
We may infer from the known results for $N=1,2$ as
\begin{eqnarray}
  \label{eq:cn}
  c_n=\Biggl(-\frac{\eta}{8\pi g\sin2\pi g}\Biggr)^n~.
\end{eqnarray}
This completes the construction of the basic operator (\ref{av}).

Exponential operator (\ref{expkappa}) for an arbitrary negative 
integer $\kappa$ can be obtained inductively from (\ref{av}) as composite 
operators \cite{tikf} 
\begin{eqnarray}
  \label{cmpop}
  e^{\kappa\eta\lambda^1\cdot\varphi(x)}=\lim_{x'\rightarrow x}
  |(1-e^{-i(x^+-x'{}^+)})(1-e^{-i(x^--x'{}^-)})|^{\Delta_{\kappa}
    -\Delta_{-1}-\Delta_{\kappa+1}}{\rm e}^{-\eta\lambda^1
    \cdot\varphi(x)}{\rm e}^{\eta(\kappa+1)\lambda^1\cdot\varphi(x')}~,
\end{eqnarray}
where $\Delta_\nu$ is the conformal weight of the free field vertex 
operator $V_\nu^1$ defined by (\ref{cdim}). In the classical theory, 
the Toda exponential for $\kappa $ being a negative integer is a finite 
polynomial of the screening charges as can be seen from (\ref{clexp}). 
This is expected to hold also in quantum theory. Our construction is 
consistent with this.

Using (\ref{expkappa}) and 
(\ref{av}) in (\ref{cmpop}), we obtain the recurrence relation for 
$C^1_{n_1\cdots n_N}(\kappa;\varphi)$ as
\begin{eqnarray}
C^1_{n_1\cdots n_N}(\kappa;\varpi)
=\sum_{m=0}^{N}C^1_{[m]}(\varpi-\kappa\lambda^1)
C^1_{n_1\cdots n_{m-1} n_m -1 n_{m+1}\cdots n_N}(\kappa +1;\varpi
+\lambda^1-\mu_m)~.
\label{receq}
\end{eqnarray}
In deriving this, use has been made of 
\begin{eqnarray}
  \label{eq:uf}
V_{\nu}^1(x)\varpi=(\varpi +\nu\lambda^1)V_{\nu}^1(x)~,\qquad
A_{\{m\}}(x)\varpi=(\varpi+\lambda^1-\mu_m)A_{\{m\}}(x)~.  
\end{eqnarray}
The recurrence relation can be analyzed inductively with respect to $-\kappa$ and 
$N$, the rank of Toda theory. We finally arrive at the closed form of the expansion 
coefficients
\begin{eqnarray}
C^1_{n_1\cdots n_N}(\kappa;\varpi)
&=&(-1)^{\bar n}
\Biggl(-\frac{\eta}{8\pi g\sin2\pi g}\Biggr)^{\sum_{k=1}^N kn_k}
\frac{[\kappa]_{\bar n}}{%
  [n_1]!\cdots[n_N]!} \nonumber\\
&&\times\prod_{k=1}^N\prod_{l=0}^{k-1}
\frac{1}{[\Omega_k-\Omega_l+\bar n\delta_{l0}-n_l]_{n_k}
[\Omega_k-\Omega_l-\kappa\delta_{l0}+1]_{n_k}} 
\label{finalsolution}
\end{eqnarray}
where $n_0\equiv0$ and $\bar n\equiv n_1+\cdots+n_N$. 

So far we have 
considered $\kappa$ as arbitrary negative integers. We may continue 
$\kappa$ to arbitrary real numbers. It is almost obvious that 
the Toda exponential operator (\ref{expkappa}) with (\ref{finalsolution}) 
for arbitrary real $\kappa$ automatically satisfy the locality condition. 
Furthermore, the exact operator solutions for the Liouville and $A_2$ cases 
are also reproduced in this way. 

We have investigated quantum $A_N$-Toda field theory and, in particular, 
a construction of the exponential operator associated with the fundamental 
weight vector $\lambda^1$. Though this is only a first step toward the 
full operator solution, our result includes the operator solutions 
for the Liouville and the $A_2$ cases, where only the screening charge
operators of the type given by (\ref{quana1}) appear. For $N\geq3$ we 
encounter screening charges other than $A_{\{n\}}$. For instance 
in the $A_3$-system we can give the exponential operators associated 
with the fundamental weights $\lambda^1$ and $\lambda^3$. To construct 
$e^{\kappa\eta\lambda^2\cdot\varphi}$ we must seek for the quantum 
expressions of the screening charges $A_2$, $A_{21}$, $A_{23}$, 
$A_{213}+A_{231}$, $A_{2132}+A_{2312}$ and then analyze their 
exchange algebra. (See \cite{tikf} for notation.) 
These are not of the type $A_{\{n\}}$ and are expected to have 
rather complicated exchange algebra. We will investigate them 
in future publication.

\end{document}